# Self-amplification of radiation from an electron bunch inside a waveguide filled with periodic medium

A.R. Mkrtchyan, L.Sh. Grigoryan[*], A.A. Saharian, A.H. Mkrtchyan,
H.F. Khachatryan and V.Kh. Kotanjyan

*Institute of Applied Problems of Physics NAS RA,*
*25 Hr. Nersessian Str., 0014 Yerevan, Republic of Armenia*
*E-mail*: `levonshg@mail.ru`

ABSTRACT: We investigate the radiation from a bunch of relativistic electrons moving along the cylindrical waveguide axis, assuming that the waveguide is partially loaded by a medium with periodic dielectric permittivity and magnetic permeability. The spectral distribution of the radiation energy flux through the cross section of the waveguide is studied at large distances from the medium. The analysis is based on the corresponding exact solution of Maxwell equations for the case of a single electron moving along the waveguide axis. The results of numerical calculations are presented in the special case of layered medium consisting of a finite number of dielectric plates separated by vacuum gaps. We show that under certain conditions on the problem parameters the quasi-coherent Cherenkov radiation generated by the electron bunch inside the plates is self-amplified at certain waveguide modes. A visual explanation of this phenomenon is provided that reproduces the main features to rather good accuracy.

KEYWORDS: Cherenkov radiation; Waveguide; Stack of plates; Bunch of relativistic electrons.

---

[*] Corresponding author.

# Contents

## 1. Introduction

It is well known that the presence of medium may essentially influence the characteristics of the radiation from charged particles. Moreover, the medium gives rise to new types of the radiation processes (see, for example, [1-5]). Examples of the latter are the Cherenkov radiation (CR), transition radiation, diffraction radiation etc. These processes are used as sources for the electromagnetic radiation in wide range of frequencies, in particle detectors and also for radiation source diagnostics. In addition, by a comparison of the measured spectral and angular characteristics of the radiation with the theoretical ones the relevant parameters of the medium can be extracted. New interesting effects in the radiation processes arise in periodic media. Effects of periodic structures on the coherence properties of the radiation provide an additional tool for the control of the radiation characteristics.

    In the present paper we consider the CR from a bunch of charged particles in a cylindrical waveguide partly loaded by a periodic medium. The CR in waveguides with dielectric filling has been widely discussed in the literature (for an early review see [1] and references given in [6]). In particular, special attention has been paid to the radiation on periodical structures inside the waveguide (see, for example, [7-12] and references therein). In [7] the radiation from a charged particle is considered inside an infinite waveguide completely filled with a layered (spatially periodic) medium. However, the case of the CR generation was not discussed in that reference and it has been investigated in [9,10]. It was shown that CR may self-amplify due to the presence of a waveguide and of a periodic medium.

    The organization of the present paper is as follows. In the next section we formulate the problem and consider the radiation from a single particle. The numerical results and the features of the radiation are discussed in section 3. Section 4 is devoted to the visual explanation of the radiation enhancement for a special choice of the problem parameters. The main results are summarized in section 5.

## 2. Problem setup

Consider a thin bunch of relativistic particles with charge $q$ uniformly moving with velocity v along the axis of an infinitely long cylindrical waveguide with perfectly conducting walls. The radius of the waveguide will be denoted by $R$ and cylindrical coordinates $r, \varphi, z$ with the axis $z$ along the waveguide axis will be used. We shall assume that the finite part of the waveguide is loaded by a laminated medium with weak absorption of the radiation. An example of a waveguide loaded by a stack of plates is shown in figure 1 (for the electromagnetic field of the bunch intersecting a dielectric plate in a waveguide, or a vacuum gap located in a dielectric loaded waveguide see [13, 14]).

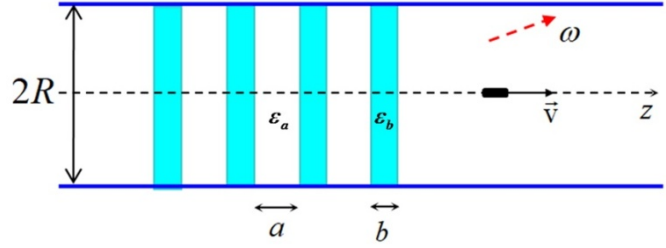

**Figure 1.** Geometry of the problem. The period of layered medium is $l = a + b$.

Let the permittivity $\varepsilon$ and permeability $\mu$ of the medium are independent of the transverse coordinates $r, \varphi$, and are general periodic functions of the axial coordinate (within the region of the medium occupation) with the period $l$: $\varepsilon(z+l) = \varepsilon(z)$, $\mu(z+l) = \mu(z)$. We denote by $I_n(\omega)$ the spectral density of the radiation energy on the $n$-th mode of the waveguide, passing the waveguide cross section (during the whole period of the bunch motion) at large distances from the medium and corresponding to the radiation from a single particle. Then, for the corresponding energy of the radiation emitted by the bunch we can write

$$W^* = \sum_{n=1}^{\infty} \int_{\omega_n}^{\infty} F(\omega) I_n(\omega) d\omega = \sum_n W_n^* \qquad (2.1)$$

where $\omega_n$ is the cutoff frequency for the $n$-th mode and $W_n^*$ is the energy radiated on that mode. In (2.1) the factor $F(\omega)$ is determined by the bunch structure and is presented as [15]

$$F(\omega) = n_q[1 - f_q(\omega)] + n_q^2 f_q(\omega), \qquad (2.2)$$

with $n_q$ being the number of particles in the bunch. For the Gaussian distribution of the bunch particles with standard deviation $\sigma$ the coherence factor is given by $f_q = \exp(-\omega^2 \sigma^2 / v^2)$.

Here we aim to specify the conditions on the parameters of the problem under which the combined effects of the waveguide and of the periodical structure of the loaded medium strongly influence the intensity and the spectral distribution of the radiation from a bunch of particles when the Cherenkov condition is satisfied. We also provide an explanatory visualization of the effect.

First let us consider the case of a single particle uniformly moving along the waveguide axis with the velocity v ($z = vt$). By taking into account the azimuthal symmetry of the problem and making the Fourier transform $f_\omega = \int f(t)\exp(i\omega t)dt/2\pi$, one may reduce the Maxwell equations to the single equation [7]

$$\left[\varepsilon\frac{d}{dz}\left(\frac{1}{\varepsilon}\frac{d}{dz}\right)+\frac{\omega^2}{c^2}\varepsilon\mu-\frac{\alpha_n^2}{R^2}\right]A_n = \frac{\varepsilon}{v}\frac{d}{dz}\left(\frac{1}{\varepsilon}e^{i\omega z/v}\right)-i\frac{\omega}{c^2}\varepsilon\mu e^{i\omega z/v}, \qquad (2.3)$$

for the function $A_n = A_n(z)$. The longitudinal component of the electric field, $E_{z\omega}$, is expressed in terms of this function by means of formula

$$\varepsilon E_{z\omega}(r,z) = \sum_{n=1}^{\infty}\frac{2q}{\pi R^2 J_1^2(\alpha_n)}J_0(\tfrac{\alpha_n}{R}r)A_n(z). \qquad (2.4)$$

Here and in what follows $J_\nu(x)$ is the Bessel function of the first kind, $\alpha_n$ is the $n$-th zero of the function $J_0(x)$: $J_0(\alpha_n) = 0$, and the coefficient $2q/\pi R^2 J_1^2(\alpha_n)$ is extracted for the further convenience. In formulas (2.3) and (2.4), $\varepsilon \equiv \varepsilon_\omega$ and $\mu \equiv \mu_\omega$ are the Fourier transforms of the permittivity and permeability. In the problem with a single particle all the features can be obtained on the base of the equation (2.3).

For waves propagating in the hollow part of the waveguide (with $\varepsilon = \mu = 1$) one has

$$A_n(z) = A_n^q(z) + \frac{i}{\omega}a_2\exp(ik_2 z), \qquad k_2 = \sqrt{\omega^2/c^2 - \alpha_n^2/R^2}. \qquad (2.5)$$

Here the first term in the right-hand side corresponds to the known field of the charge inside the hollow waveguide, and the second term (with dimensionless factor $a_2$) describes the free field (the radiation field), if the condition $\omega > \omega_n = \alpha_n c/R$ is obeyed.

By using the Poynting vector for the radiation field at large distances from the medium, described by the second term in the right-hand side of (2.5), for a single particle the energy flux through the waveguide cross-section is presented as

$$W = \int_{-\infty}^{\infty}dt\int_S d\sigma \frac{c}{4\pi}(\vec{E}\times\vec{H})\cdot\vec{e}_z = \sum_n\int_{\omega_n}^{\infty}I_n(\omega)d\omega, \qquad (2.6)$$

where $S$ stands for the cross-section. The spectral distribution of the radiation, introduced in (2.1), is determined by the formula [10]

$$I_n(\omega) = \frac{4q^2}{\pi\omega}\frac{k_2|a_2|^2}{\alpha_n^2 J_1^2(\alpha_n)}. \qquad (2.7)$$

For the evaluation of the amplitude $a_2$ one needs to compare (2.5) with the solution of the equation (2.3) valid for all $-\infty < z < \infty$. Such a solution can be obtained by using the method of the Green functions [10].

The formulas (2.3)-(2.7) are general and they are valid for any functions $\varepsilon(z), \mu(z)$ describing a finite medium inside the waveguide (including the case of aperiodic functions). The specific form of the functions $\varepsilon(z)$, $\mu(z)$ is required for the evaluation of the coefficient $a_2$. Below we shall consider a special, but highly advantageous case of a laminated medium inside the waveguide consisting $N$ plates with the thickness $b$ separated by vacuum gaps having thickness $a$ ($\varepsilon_a = \mu_a = 1$ in figure 1, see also [8] and references therein). In this special case the coefficient $a_2$ can be evaluated in different ways. One of the methods could be to present the field in separate regions with constant permittivity and permeability as a superposition of two solutions (plane waves) propagating to the right and to the left. The corresponding coefficients are determined by the boundary conditions on interfaces separating the layers. For the case of periodic $\varepsilon(z)$, $\mu(z)$ and for a medium occupying a finite region we can consider another method. Locally, inside the medium the Maxwell equations are the same as those for an infinite medium with the same $\varepsilon(z)$ and $\mu(z)$. Hence, the two independent solutions of (2.3) for finite and infinite media with the same $\varepsilon(z)$, $\mu(z)$ are the same functions (Bloch functions). The corresponding solutions differ in the coefficients when forming the superposition of two independent solutions. In the problem at hand, those coefficients are determined by using the boundary conditions on the first and last interfaces of the finite layered medium. The boundary conditions on the remaining interfaces are automatically included in the Bloch solutions. The fields evaluated in this way contain all the information about the radiation features, in particular, the ones corresponding to the transition radiation.

Using the results obtained in [10] by the second method described above, for a finite layered medium the amplitude $a_2$ can be presented in the form

$$a_2 = \frac{\omega}{vk_a A_-} \left( \frac{\psi_a}{2} \left( \frac{A_+}{1+vk_a/\omega} + \frac{A_- - 2B/q_+^N}{1-vk_a/\omega} \right) - Y_1 w_1 + Y_2 w_2 \right), \qquad (2.8)$$

where and in what follows $\psi_j = 1/\varepsilon_j - \mu_j \cdot v^2/c^2$, $k_j = (\alpha_n/R) \cdot \sqrt{\varepsilon_j \mu_j \omega^2/\omega_n^2 - 1}$, $j = a, b$,

$$w_1 = e^{i\frac{\omega a}{v 2}} \cdot \left( \psi_a \left( \left( e^{-i(\frac{\omega}{v}-k_a)\frac{a}{2}} + \gamma_1 e^{i\frac{\omega}{v}b} \right) \cdot \frac{e^{i(\frac{\omega}{v}-k_a)\frac{a}{2}}-1}{1-v\cdot k_a/\omega} + \delta_1 e^{i\frac{\omega}{v}b} \cdot \frac{e^{i(\frac{\omega}{v}+k_a)\frac{a}{2}}-1}{1+v\cdot k_a/\omega} \right) + \right.$$
$$\left. +\psi_b \left( \alpha_1 \frac{e^{i(\frac{\omega}{v}-k_b)b}-1}{1-v\cdot k_b/\omega} + \beta_1 \frac{e^{i(\frac{\omega}{v}+k_b)b}-1}{1+v\cdot k_b/\omega} \right) + \left( \frac{1}{\varepsilon_b} - 1 \right) \left( 1 - e^{i\frac{\omega}{v}b} \left( \alpha_1 e^{-ik_b b} + \beta_1 e^{ik_b b} \right) \right) \right),$$

$$w_2 = e^{i\frac{\omega a}{v 2}} \cdot \left( \psi_a \left( \left( e^{i\frac{\omega}{v}b} + \gamma_1 e^{-i(\frac{\omega}{v}+k_a)\frac{a}{2}} \right) \cdot \frac{e^{i(\frac{\omega}{v}+k_a)\frac{a}{2}}-1}{1+v\cdot k_a/\omega} + \delta_1 e^{-i(\frac{\omega}{v}-k_a)\frac{a}{2}} \cdot \frac{e^{i(\frac{\omega}{v}-k_a)\frac{a}{2}}-1}{1-v\cdot k_a/\omega} \right) + \right.$$
$$\left. +\psi_b \left( \alpha_1 \cdot \frac{e^{i\frac{\omega}{v}b}-e^{-ik_b b}}{1+v\cdot k_b/\omega} + \beta_1 \cdot \frac{e^{i\frac{\omega}{v}b}-e^{ik_b b}}{1-v\cdot k_b/\omega} \right) + \left( \frac{1}{\varepsilon_b} - 1 \right) \left( \alpha_1 e^{-ik_b b} + \beta_1 e^{ik_b b} - e^{i\frac{\omega}{v}b} \right) \right),$$

$A_+ = L_{1+} L_{2+} (1 - 1/Q^{2N})$, $A_- = L_{1-} L_{2+} - L_{1+} L_{2-}/Q^{2N}$, $B = L_1 L_{2+} - L_{1+} L_2$,

$$Y_m = p_{m;1} L_{2+} \frac{1-1/q_+^N}{q_+ - 1} - p_{m;2} L_{1+} \frac{1-1/q_-^N}{Q^{2N}(q_- - 1)}, \quad q_\pm = Q^{\pm 1} e^{i\omega l/v}, \quad m = 1, 2,$$

$p_{1;1} = u_{1;0} - Q u_{2;0}$, $p_{1;2} = u_{1;0} - Q^{-1} u_{2;0}$, $p_{2;1} = u_{2;0} - Q u_{1;0}$, $p_{2;2} = u_{2;0} - Q^{-1} u_{1;0}$,

$$L_m = p_{1;m} u_{1;0} - p_{2;m} u_{2;0}, \quad L_{m\pm} = p_{1;m} u_{1\pm} - p_{2;m} u_{2\pm},$$

$u_{1+} = 0$, $u_{1-} = 2u_{1;0} = 2e^{ik_a a/2}$, $u_{2+} = 2\gamma_1 e^{-ik_a a/2}$, $u_{2-} = 2\delta_1 e^{ik_a a/2}$, $u_{2;0} = (u_{2+} + u_{2-})/2$,

$\delta_1 = 0.5i(\sigma_* - 1/\sigma_*)\sin(k_b b)$, $\gamma_1 = \cos(k_b b) - 0.5i(\sigma_* + 1/\sigma_*)\sin(k_b b)$, $\sigma_* = k_b/(\varepsilon_b k_a)$,

$$\alpha_1 = 0.5(1+1/\sigma_*), \quad \beta_1 = 0.5(1-1/\sigma_*). \qquad (2.9)$$

In the expressions above, $Q$ is defined by the relations

$$Q + 1/Q = 2\cos(k_b b)\cos(k_a a) - (\sigma_* + 1/\sigma_*)\sin(k_b b)\sin(k_a a), \quad |Q| \geq 1. \qquad (2.10)$$

## 3. Numerical results

The CR generated by a single electron inside a waveguide loaded by a stack of plates is investigated in [10]. In the present paper the CR from a bunch of electrons is studied and quantitative analysis for estimation of the amplification or suppression coefficient of the generated radiation is provided.

**Table 1.** Values of the parameters for the graphs in figure 2.

| Curve | A | B | C | D | G |
|---|---|---|---|---|---|
| $a/R$ | 200 | - | - | 15.43 | - |
| $b/R$ | 13.79 | 4×13.79 | 13.79 | 13.79 | 4×13.79+3×15.43 |

In figure 2 we display three curves $A,B,C$ (the curves $D$ and $G$ in figure 2 will be discused below) for the spectral distribution $F(\omega)I_n(\omega)$ of the radiation energy from a bunch of electrons on the 3$^{rd}$ mode of the waveguide. The graphs are plotted for the values of the parameters given in table 1 and for the waveguide radius we have taken $R = 1\,\text{cm}$. The permittivity and permeability for the material of the plates are given by $\varepsilon_b = \varepsilon'_b + i\varepsilon''_b = 1.3 + 0.002i$ and $\mu_b = 1$. The energy of electron is 1.2 MeV so that the Cherenkov condition is satisfied: $\text{v} > c/\varepsilon'^{1/2}_b$. For the remaining parameters we have taken $\sigma = 0.02\,\text{cm} << 2\pi c/\omega$ and $n_q = 10^9$.

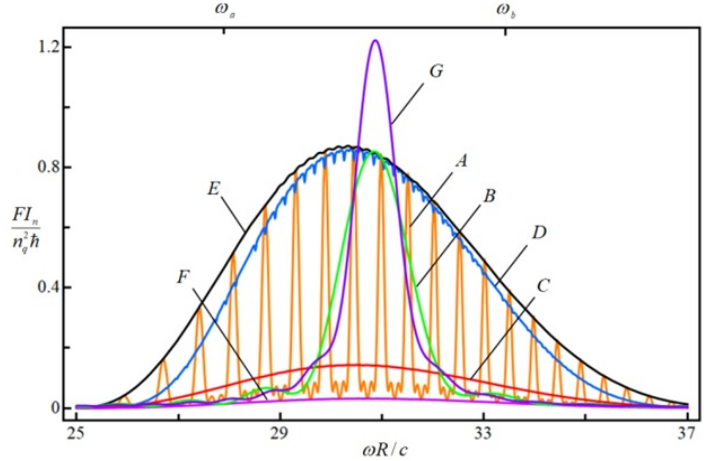

**Figure 2.** The spectral distribution of the Cherenkov radiation energy from an electron bunch on the 3$^{rd}$ mode of waveguide. For details see the text and table 1.

The curve $A$ in figure 2 corresponds to the case when the waveguide is loaded with a stack of $N = 4$ plates (column A in table 1). The oscillatory nature of that curve is caused by the superposition of the CR pulses generated by the bunch in separate plates. The curve $B$ describes the radiation in the case when the four plates form a single homogeneous thick plate by removing the vacuum gaps (column B in table 1). And finally, the curve $C$ presents the spectral distribution of the radiated energy for a single plate with thickness $b$ (column C in table 1). As we could expect, for the cases B and C the oscillations in the spectral distribution of the radiated energy are absent.

Now consider the energy of the radiation on the 3$^{rd}$ mode of the waveguide in the frequency range $(\omega_a, \omega_b)$ corresponding to the half-width of the curve $C$:

$$W_3 = \int_{\omega_a}^{\omega_b} F(\omega) I_3(\omega) d\omega, \quad I_3(\omega_a) = I_3(\omega_b) = I_3(\omega_{max})/2, \quad (3.1)$$

where

$$\frac{\omega_a}{2\pi} = \frac{27.89c}{2\pi R} = 0.133\,\text{THz}, \quad \frac{\omega_b}{2\pi} = \frac{33.43c}{2\pi R} = 0.16\,\text{THz}. \quad (3.2)$$

By taking into account that, in general, the total energy in the interference process is not changed, we could expect that the radiation energy for the cases A and B should not differ essentially: $W_3^A = 3.61 \cdot 10^{-6}\,\text{J}$, $W_3^B = 4.64 \cdot 10^{-6}\,\text{J}$. Note that in the case of a single plate one has $W_3^C = 2.05 \cdot 10^{-6}\,\text{J}$. It should also be noted that the location of the maximum in the spectral distribution of the CR energy is given by the known simple formula: $\omega_{max} \approx \alpha_n \text{v}/(R\sqrt{\varepsilon_b' \text{v}^2/c^2 - 1})$ (see, e.g., [1]). All these features were expected and it seems that the situation is rather clear. However, it is interesting that another situation is possible.

We return to the case of a stack of four plates and select another special value of the parameter $a$ corresponding to $a/R \approx 15.43$ (table 1, column D, for the choice of this specific value see the next section). The corresponding spectral distribution of the radiated energy is depicted in the same figure 2 (curve $D$). It is seen that compared to the case A the oscillations practically disappear. Moreover, the energy of the radiation (area under the curve $D$) in the frequency range $(\omega_a, \omega_b)$ is about three times larger than that for the curves $A$ or $B$: $W_3^D = 1.17 \cdot 10^{-5}\,\text{J}$, and is about six times larger than that for curve $C$: $W_3^D/W_3^C \approx 5.71$. The latter factor is larger than the number of the plates. It is of interest to provide a visual explanation for such essential enhancement of the radiation energy.

## 4. Visual explanation

Below we provide a simple model that reproduces the basic features of the numerical results. In what follows, a bunch of particles will be considered as a point charge (charged particle) to simplify the presentation.

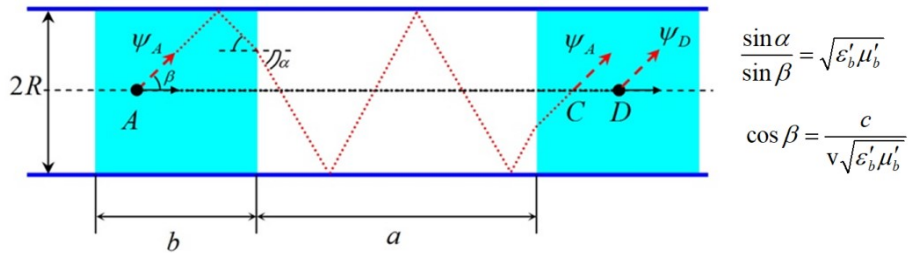

**Figure 3.** Two neighboring plates inside the waveguide.

Consider the instant when the particle is in the left plate (see figure 3) and follow the pulse $\Psi_A$ of the CR generated by the particle near the point A (dashed line). Propagating inside the waveguide the pulse enters the right plate and crosses the waveguide axis near the particle trajectory at certain point C. By that time the particle, being at the point D, emits another pulse of the CR (shown in figure 3). If the conditions , are satisfied, presented as

$$\frac{a+b}{v} = \frac{a}{c\cos\alpha} + \frac{b\sqrt{\varepsilon'_b \mu'_b}}{c\cos\beta}, \qquad \frac{a\tan\alpha + b\tan\beta}{2R} \approx 4s \qquad (4.1)$$

with $s$ being an integer, the point $D$ coincides with the point $C$ and the relative location of the point D in the right plate is the same as that for the point $A$ in the left plate.

As a result of interference, depending on the specific value of the phase difference, the superposed pulses $\Psi_A$ and $\Psi_D$ can be either amplified or reduced. The factor 4 in the right-hand side of the second relation in (4.1) ensures the fulfillment of the condition for constructive superposition of the pulses (see, e.g., [2]). The thicknesses of the vacuum gaps and the plates for the curve $D$ in figure 2 (see table 1, the column D) have been chosen on the base (4.1) with $s=1$.

The in-phase superposition of waves near the point D gives rise to the total field increase in the radiation formation zone. This explains the physical reason for the CR amplification (more precisely, self-amplification). In what follows we present a simplified numerical analysis that reproduces the self-amplification phenomenon to rather good accuracy. Let us denote by $E = E(\omega)$ the spectral component of the electric field strength projection on the waveguide axis corresponding to the CR by the particle inside a single plate in waveguide (considered near the point of the particle location). For the corresponding field acting on the particle inside the $k$-th plate we can write the following simple relation

$$E_k(\omega) = \begin{cases} 0.5E + \eta E + \eta^2 E + \ldots + \eta^{k-1} E = E \cdot (0.5 + \sum_{n=1}^{k-1}\eta^n) & \text{for } k>1 \\ 0.5E & \text{for } k=1 \end{cases} \qquad (4.2)$$

or, after summation,

$$E_k(\omega) = \frac{(1+\eta)/2 - \eta^k}{1-\eta} E. \qquad (4.3)$$

It consists of the contribution from the radiation field $0.5E(\omega)$ generated inside that plate (the factor 0.5 takes into account the fact that the particle moves while remaining at the front of the generated CR wave), and the radiation fields generated in the previous plates of the stack. The coefficient $\eta$ takes into account the CR wave attenuation after passing through the plate:

$$(1-r)^2 e^{-b/l_{Ch}} \equiv \eta^2, \qquad r = \frac{\tan^2(\alpha-\beta)}{\tan^2(\alpha+\beta)}, \qquad (4.4)$$

where $r$ is the coefficient of reflection of the CR from each of two boundaries of the plate. In (4.4), $l_{Ch}$ is the distance along the direction of particle motion, after which the energy of the CR decreases by the factor of $e$.

It is also clear that the spectral density of the CR energy, $J^{(i)}(\omega)$, generated in $i$-th plate, decreases after passing through the next ($i$+1)-th plate and becomes equal $\eta^2 J^{(i)}(\omega)$. For this reason, the spectral density for the total radiation energy, $I^{(N)}(\omega)$, will be determined by the following expression

$$I^{(N)}(\omega) = J^{(N)}(\omega) + \eta^2 J^{(N-1)}(\omega) + \eta^{2\cdot 2} J^{(N-2)}(\omega) + \ldots + \eta^{2(N-1)} J^{(1)}(\omega). \qquad (4.5)$$

$$I^{(N)}(\omega) \sim E \cdot \sum_{j=1}^{N} \eta^{2(N-j)} (0.5 + (1-\delta_{j1}) \sum_{n=1}^{j-1} \eta^n) = 0.5 E \cdot \left( \frac{1-\eta^N}{1-\eta} \right)^2, \quad I^{(N)}(\omega) \cong \xi_N^+(\omega) J^{(1)}(\omega), \quad (4.6)$$

where $\xi_N^+(\omega) \cong (1-\eta^N)^2 /(1-\eta)^2$. If the dependence of $\eta$ on the frequency $\omega$ is week then the same is the case for $\xi_N^+(\omega)$. In that case for the radiated energy one obtains $W^D \cong \xi_N^+(\omega_{max}) W^C$. Equations (4.2), (4.3) are valid for the constructive superposition of waves of the CR (self-amplification mode). In the case of the destructive interference of the CR waves in the considerations given above one needs to make the replacement $\eta \to -\eta$. In a similar way we can see that $I^{(N)}(\omega) \cong \xi_N^-(\omega) J^{(1)}(\omega)$, where $\xi_N^-(\omega) \cong (1-(-\eta)^N)^2 /(1+\eta)^2$. For the values of the parameters presented above one gets $\xi_N^+(\omega_{max}) = 5.99$, $\xi_N^-(\omega_{max}) = 0.22$.

Similar relations are obtained for the maxima and minima $I^{(N)}(\omega_\pm)$ in the spectral distribution of the energy radiated in plates, radiating independently from each other. Indeed, for the maxima and minima of the electric field strength inside the waveguide at large distances from the plates one has $E_N(\omega_\pm, \infty) = E_1 \pm \eta E_1 + \eta^2 E_1 \pm \ldots + (\pm \eta)^{N-1} E_1$. By taking into account that $I^{(N)}(\omega_\pm) \sim E_N^2(\omega_\pm, \infty)$, we get

$$I^{(N)}(\omega_\pm) \cong \xi_N^\pm J^{(1)}(\omega_\pm), \qquad \xi_N^\pm = \left[ \frac{1-(\pm\eta)^N}{1 \mp \eta} \right]^2. \quad (4.7)$$

However, as it already has been mentioned, the total energy $W_N^*$ radiated by the plates independently from each other is not changed when changing their relative positions (interference).

In figure 2 the curves $E$ and $F$ describe the spectral distributions of the radiated energy from a single plate (corresponding to the curve $C$) multiplied by the self-amplification $\xi_N^+(\omega)$ (curve $E$) or self-suppression $\xi_N^-(\omega)$ (curve $F$) coefficients. To great accuracy, the maxima of the curve $A$ are located near the curve $E$, and the minima are near the curve $F$, that confirms the correctness of our visual consideration. Moreover, (i) the shift between the curves $D$ and $E$ is small near the maximum and (ii) the ratio of radiation energies in the cases $D$ and $C$ is equal to $W_3^D / W_3^C \approx 5.71 \approx \xi_N^+(\omega_{max}) = 5.99$.

The curve $G$ in figure 2 (see also table 1, the column G) corresponds to the spectral distribution of the radiation energy generated inside a single thick plate with a thickness $4b + 3a$. The curve $D$ presents the same quantity for a stack of plates, with the same thickness. Surprisingly, $W_3^D$ exceeds $W_3^G$ more than two times: $W_3^D / W_3^G = 2.29$. This means that by removing a part of the matter from a thick plate, one can increase the radiation energy more than two times. Similar results are obtained in the case when the waveguide is loaded by two plates.

The features discussed above are determined by the ratios $a/R$ and $b/R$. For given values of these ratios they are not sensitive to the absolute value of the waveguide radius. Note that related to terahertz radiation applications and electron bunch acceleration and compression in charged particle accelerators, the dielectric loaded waveguides have been widely discussed in the literature with radii ranging from sub-millimeter to several centimeters. For example, sub-millimeter radius waveguides were used in [16,17], where terahertz coherent Cherenkov radiation has been observed from cylindrical dielectric-lined waveguides. We have demonstrated that the periodic structure inside the waveguide serves as an additional tool to control the radiation parameters. In particular, the radiation intensity can be amplified by tuning the parameters in accordance with (4.1). The velocity of the particle enters in that condition in the form of the ratio $v/c$ and for sufficiently high energies its dependence on the particle energy is weak. This shows that the amplification effect is not critically sensitive to the particle energy. For example, instead of the electron energy $1.2\,\text{MeV}$, used in numerical evaluations above, we could take the energy $10\,\text{MeV}$ considered in the experiments of [16].

## 5. Summary

We have investigated the CR from a bunch of charged particles uniformly moving along the axis of a cylindrical waveguide loaded by a stack of dielectric plates with vacuum gaps between them. It is shown that by tuning the parameters of the system the energy of the CR can be increased by several times compared to the radiation from a single plate of the same material obtained by removing the vacuum gaps. We have provided a simple visual explanation for this kind of self-amplification effect that reproduces the main features of the phenomena to rather good accuracy. The effect considered can be used to develop high-power sources of the coherent CR in giga and terahertz spectral ranges, e.g., for the amplification of the coherent CR observed in [16,17]. Though, we have considered the effect on the example of the stack of plates, we expect that the features discussed should take place also in other periodic structures.

## Acknowledgments

The authors are grateful to X. Artru, A.V. Tyukhtin and A. Tishchenko for discussions and comments. The work was supported by the RA Committee of Science, in the frames of the research project №18T-1C397.